\documentclass[12pt]{article}
\newcommand{\be}{\begin{equation}}
\newcommand{\ee}{\end{equation}}
\newcommand{\bea}{\begin{eqnarray}}
\newcommand{\eea}{\end{eqnarray}}
\begin{document}
\title{Proper magnetic fields for nonadiabatic geometric quantum gates in NMR
  }
\author{Kazuto Oshima\thanks{E-mail: oshima@nat.gunma-ct.ac.jp} and Koji Azuma   \\ \\
\sl Gunma National College of Technology, Maebashi 371-8530, Japan }

\date{}
\maketitle
\begin{abstract}
In a scheme of nonadiabatic purely geometric quantum gates in nuclear magnetic resonance(NMR) systems we propose proper magnitudes of magnetic fields that are suitable for an  experiment. We impose a natural condition and reduce the degree of freedom of the magnetic fields to the extent.  By varying the magnetic fields with essentially one-dimensional degree of freedom, any spin state can acquire arbitrary purely geometric phase
 $\phi_{g}=-2\pi(1-\cos{\theta})$, $0 \le \cos{\theta} \le 1$.  This is an essential ingredient for constructing universal geometric quantum gates.  
\end{abstract}

\newpage
The geometric phase is expected to be useful in constructing fault tolerant quantum logical gates
for quantum computation. 
An experiment of quantum gates based on 
the adiabatic geometric phase \cite{Berry} has been done \cite{Jones} in NMR systems.
The adiabatic quantum computation has the disadvantage that
we cannot reduce the gate operation time because of the adiabatic condition.
Accordingly, it is very hard to preserve the coherency in the adiabatic computation.
Wang and Keiji \cite{Matsumoto} have proposed quantum gates based on the nonadiabatic geometric phase \cite{Anandan}.
In their scheme the gate operation time can be varied by the angular velocity
of an external rotating magnetic field.   However, as the gate operation time
becomes short the dynamic phase grows.  The dynamic phase
is unwelcome, because it is thought to be easily affected by quantum noises.
One of the ways to construct nonadiabatic purely geometric gates is 
to add such a magnetic field as a spin develops to the {\it horizontal}
direction \cite{horizontal} where the dynamic phase does not accumulate \cite{Matsumoto}.
Another way is to develop the spin along plural loops on the Bloch sphere so
that the  corresponding dynamic phases are canceled among them \cite{Matsumoto,Zhu1}.  

Recently, Zhu and Wang \cite{Zhu2}
have studied the multi-loop method and shown that under suitable
external rotating magnetic fields the dynamic phases corresponding to  
two loops on the Bloch sphere are canceled each other.
However, there leaves residual freedom in the magnitudes of the magnetic fields
and it is inevitable to reduce the degree of freedom to the extent
for implementation of an efficient NMR experiment.   The purpose of
this paper is to propose a suitable condition to reduce the degree
of freedom and exhibit proper magnitudes of the magnetic
fields for an actual NMR experiment.   Their scheme is based on the fact that universal quantum gates can be constructed
from two noncommutable single-qubit gates and one nontrivial two-qubit gate \cite{Lloyd}.   For simplicity,  we confine ourselves to single-qubit gates
in this paper. 

Let us start from the spin up state ${\psi}={}^{t}(1,0)$.  This state can nonadiabatically be transformed by the $S$ operation in Ref.\cite{Matsumoto} to a state
$\psi_{+}={}^{t}(\cos{\theta \over 2},\sin{\theta \over 2})$ and vice versa.   This state and an 
orthogonal state $\psi_{-}={}^{t}(-\sin{\theta \over 2},\cos{\theta \over 2})$ can be used as a pair of 
cyclic states, which evolve to 
the starting states up to phases under a periodic process.
We only consider the state $\psi_{+}$, because the phase $\psi_{-}$
acquires is opposite to the phase $\psi_{+}$ acquires \cite{Zhu3}.   Moreover it will be
suffice to confine ourselves to the case $0 \le \theta \le {\pi/2}$,
because the geometric phase we get in the following is $\phi_{g}=-2\pi(1-\cos{\theta})$.
The Hamiltonian $H$ of a nuclear-spin under a magnetic field ${\bf B}$ is
\begin{equation}
 H=-{\gamma \over 2}{\bf B}\cdot{\vec{\sigma}},
\label{hamiltonian}
\end{equation}
where ${\gamma}$ is the nuclear magnetomechanical ratio. The schr${\rm{\ddot o}}$dinger equation for a rotational magnetic field ${\bf B}=-(\omega_{1}\cos{\omega}t,\omega_{1}\sin{\omega}t,\omega_{0})/{\gamma}$ is
\begin{equation}
i{\partial \over \partial t}\psi={1 \over 2}(\omega_{1}\sigma_{x}\cos{\omega t}+\omega_{1}\sigma_{y}\sin{\omega t}+\omega_{0}\sigma_{z}){\psi}.
\label{schrodinger}
\end{equation} 
If $\psi_{+}$ is an eigenstate of the corresponding
rotational frame Hamiltonian $H_{0}$,
\begin{equation}
H_{0}={1 \over 2}(\omega_{0}-\omega)\sigma_{z}+{1 \over 2}\omega_{1}\sigma_{x},
\end{equation}  
the following is a rotating solution of the schr${\rm{\ddot o}}$dinger equation
\begin{equation}
\psi(t)=e^{-{i \over 2}\sigma_{z}{\omega}t}e^{-iH_{0}t}\psi_{+}.
\end{equation}
At time $\tau=2\pi/{\omega}$, the unit vector ${\bf n}=\langle \psi_{+}|{\vec \sigma}|\psi_{+}\rangle$ corresponding to the state $\psi_{+}$ completes a $2\pi$ rotation on a cone on the Bloch sphere.
In the same way, choosing another suitable rotational magnetic field 
${\bf B}^{\prime}=(\omega_{1}^{\prime}\cos{\omega}t,    
\omega_{1}^{\prime}\sin{\omega}t, \omega_{0}^{\prime})/{\gamma}$, we can rotate the vector ${\bf n}$ on the same cone \cite{condition} in the identical direction; 
if $\psi_{+}$ is also an eigenstate of the corresponding
rotational frame Hamiltonian $H_{0}^{\prime}$,
\begin{equation}
H_{0}^{\prime}=-{1 \over 2}(\omega_{0}^{\prime}+\omega)\sigma_{z}-{1 \over 2}\omega_{1}^{\prime}\sigma_{x},
\end{equation}  
we have another rotating solution of the $\rm {schr{\ddot o}dinger}$ equation
\begin{equation}
\psi(t)=e^{-{i \over 2}\sigma_{z}{\omega}t}e^{-iH_{0}^{\prime}t}\psi_{+}.
\end{equation}
We rotate the state $\psi_{+}$ on the same cone twice by $\bf{B}$ and $\bf{B}^{\prime}$.
We have imposed the extra condition, the spin rotates on 
the same cone, other than the conditions in Ref.\cite{Zhu2}.
By virtue of this condition we only have
to use the $S$ operation at the start and at the end.

The condition that $\psi_{+}$ is a simultaneous eigenstate of $H_{0}$ and 
$H_{0}^{\prime}$ is
\begin{equation}
\tan{\theta}={\omega_{1} \over \omega_{0}-\omega}=
{\omega_{1}^{\prime} \over \omega_{0}^{\prime}+\omega},
\label{cone-condition}
\end{equation}
where $\theta$ is the vertical angle of the cone.  
After the first conical evolution the initial state $\psi_{+}$ acquires the geometric phase
$\phi_{g}=-\pi(1-\cos{\theta})$ and the dynamic phase $\phi_{d}=-\pi(\cos{\theta}+{\Omega \over \omega})$,
where $\Omega=\sqrt{(\omega_{0}-\omega)^{2}+\omega_{1}^{2}}$.
By the second conical evolution the initial state $\psi_{+}$ acquires the geometric phase
$\phi_{g}=-\pi(1-\cos{\theta})$ and the dynamic phase $\phi_{d}=-\pi(\cos{\theta}-{\Omega^{\prime} \over \omega})$, 
where $\Omega^{\prime}=\sqrt{(\omega_{0}^{\prime}+\omega)^{2}+\omega_{1}^{\prime 2}}.$  
The two dynamic phases are canceled each other under the condition
\begin{equation}
2\cos{\theta}={\Omega^{\prime}-\Omega \over \omega}.
\label{dynamic-zero}
\end{equation}
We can remove the dynamic phase by tuning the magnetic fields ${\bf B}$ and ${\bf B}^{\prime}$ so that $(\ref{dynamic-zero})$ holds.

Note that all of the quantities in (\ref{cone-condition}) and (\ref{dynamic-zero}) 
can be expressed by the following  four ratios $\omega_{i}/\omega,\omega_{i}^{\prime}/\omega(i=0,1)$.      
This means that we can set $\omega$ as the unit $\omega=1$, therefore we abbreviate $\omega$ for the time being.  
Moreover, introducing the new parameter $\kappa=\omega_{0}^{\prime}/\omega_{0}$, from Eq.(\ref{cone-condition})
we have
\begin{equation}
\omega_{0}-1=(1+\kappa){\omega_{1} \over \omega_{1}^{\prime}-\kappa\omega_{1}},
 \omega_{0}^{\prime}+1=(1+\kappa){\omega_{1}^{\prime} \over \omega_{1}^{\prime}-\kappa\omega_{1}}.
\label{omaga0}
\end{equation}
Eliminating $\omega_{0}$ and ${\omega_{0}}^{\prime}$ in $\cos{\theta}, \Omega$ and $\Omega^{\prime}$  we have 
\begin{equation}
\cos{\theta}={ {1+\kappa \over \omega_{1}^{\prime}-\kappa\omega_{1}}  \over 
\sqrt{\left({1+\kappa \over \omega_{1}^{\prime}-\kappa\omega_{1}}\right)^{2}+1} }, \Omega^{\prime}-\Omega=(\omega_{1}^{\prime}-\omega_{1}) \sqrt{\left({1+\kappa \over \omega_{1}^{\prime}-\kappa\omega_{1}}\right)^{2}+1} 
\label{cos}
\end{equation}
The vanishing condition (\ref{dynamic-zero}) for the dynamic phase allows us to express ${1+\kappa \over \omega_{1}^{\prime}-\kappa\omega_{1}}$
by $\omega_{1}^{\prime}-\omega_{1}$,
\begin{equation}
(\omega_{1}^{\prime}-\omega_{1})\left(\left({1+\kappa \over \omega_{1}^{\prime}-\kappa\omega_{1}}\right)^{2}+1\right)
=2{1+\kappa \over \omega_{1}^{\prime}-\kappa\omega_{1}}.
\label{second}
\end{equation}
Thus if we set $x=\omega_{1}^{\prime}-\omega_{1}$, $\cos{\theta}$ becomes a function of $x$,
\begin{equation}
\cos{\theta}={f(x) \over \sqrt{(f(x))^{2}+1}},
\label{cos-f(x)}
\end{equation}
where $f(x)=(1 \pm \sqrt{1-x^{2}})/x$.

As $x$ varies $0 \le x \le 1$, choosing a preferable sign in $f(x)$, we see that $\cos{\theta}$
varies $0 \le \cos{\theta} \le 1$.   Thus we can obtain arbitrary geometric phase
$\phi_{g}=-2\pi(1-\cos{\theta})$, $0 \le \cos{\theta} \le 1$, by tuning $x={\omega_{1}^{\prime}-\omega_{1} \over \omega}$ 
in the range $0 \le x \le 1$.   Using this geometric phase, arbitrary single qubit gate can be obtained by inclining the vertical direction
of the magnetic fields from the $z$ axis \cite{Zhu1,Zhu2}.

We have studied the magnitudes of the rotating magnetic fields in the scheme of 
the nonadiabatic purely geometric quantum gates in NMR systems.  
For efficient implementation of an   
NMR experiment we have found the proper magnitudes of the magnetic fields with the extremely reduced degree of freedom that give arbitrary phase shift.
For a certain value of the angular velocity $\omega$ we have at the beginning the four degrees of freedom, the two vertical fields $\omega_{0},\omega_{0}^{\prime}$ and two horizontal fields $\omega_{1},\omega_{1}^{\prime}$.
We have imposed the two conditions, the dynamic phase is zero and the spin rotates on the same cone twice.  The latter condition is preferable also in the point that we need the least number of the $S$ operation.
Imposing these two conditions, the two vertical fields are fixed as $\omega_{0}=\omega+f(x)\omega_{1}$ and 
$\omega_{0}^{\prime}=-\omega+f(x)\omega_{1}^{\prime}$. 
We have seen that the residual degree of the freedom 
$\omega_{1}^{\prime}-\omega_{1}$  bring about the arbitrary purely geometric phase shift $\phi_{g}=-2\pi(1-\cos{\theta})$, $0 \le \cos{\theta} \le 1$.


\begin{thebibliography}{99}
\bibitem{Berry}
        M.V.Berry, Proc.R.Soc.London Ser.A{\bf 392}, 45(1984).
\bibitem{Jones}
    J.A.Jones, V.Vedral, A.Ekert, and G.Casttagnoli, Nature{\bf 403}, 869(2000).    
\bibitem{Matsumoto}
    X.B.Wang and M.Keiji, Phys.Rev.Lett.{\bf 87}, 097901(2001);{\bf 88}, 179901(E)(2002).  
\bibitem{Anandan}
        Y.Aharanov and J.Anandan, Phys.Rev.Lett.{\bf 58}, 1593(1987).
\bibitem{horizontal}
        The word {\it horizontal} implies the horizontal lift in the Hilbert space $SU(2)=\{\psi\}$.
\bibitem{Zhu1}
       S.L.Zhu and Z.D.Wang, Phys.Rev.Lett.{\bf 89}, 097902(2002).
\bibitem{Zhu2}
       S.L.Zhu and Z.D.Wang, Phys.Rev.A{\bf 67}, 022319(2003).
\bibitem{Lloyd}
        S.Lloyd, Phys.Rev.Lett.{\bf 75}, 346(1995) 
\bibitem{Zhu3}
     S.L.Zhu and Z.D.Wang, Phys.Rev.Lett.{\bf 85}, 1076(2000);\\       S.L.Zhu,Z.D.Wang and Y.D.Zhang, Phys.Rev.B{\bf 61}, 1142(2000).
\bibitem{condition}
     Although this condition is suggested in Ref.\cite{Zhu2}, we utilize it to reduce the degree of freedom and the number of the $S$ operation. 
\end{thebibliography}
\end{document}